\documentclass{jjap3}
\usepackage{newtxtext,newtxmath}
\newcommand{\vecvar}[1]{\mbox{\boldmath$#1$}}
\newcommand{\e}{\mbox{e}}

\title{Theoretical investigation of vacancy related defects at 4H-SiC(000$\bar{1}$)/SiO$_2$ interface after wet oxidation}
\author{Mukai Tsunasaki, Tomoya Ono\thanks{E-mail: t.ono@eedept.kobe-u.ac.jp}, and Mitsuharu Uemoto\thanks{E-mail: uemoto@eedept.kobe-u.ac.jp}}
\inst{Department of Electrical and Electronic Engineering, Graduate School of Engineering, Kobe University, Nada, Kobe, 657-8501 Japan}
\abst{The stability and formation mechanism of the defects relevant to silicon and carbon vacancies at the 4H-SiC(000$\bar{1}$)/SiO$_2$ interface after wet oxidation are investigated by first-principles calculation based on the density functional theory. The difference in the total energy of the defects agrees with the experimental results concerning the dencity of defects. We found that the characteristic behaviors of the generation of defects are explained by the positions of vacancies and antisites in the SiC(000$\bar{1}$) substrate and that the formation of silicon and carbon vacancies is relevant to the generation mechanism of defects. The generation of silicon and carbon vacancies is attributed to the termination of dangling bonds by H atoms introduced by wet oxidation, resulting in generation of carbon-antisite--carbon-vacancy and divacancies defects in wet oxidation.}

\begin{document}
\maketitle

\section{Introduction}
\label{sec:Introduction}
SiC is a semiconductor with a wide bandgap, making it suitable for application to electronic devices operating at a high power and high frequency. One of the major advantages of SiC metal-oxide-semiconductor field-effect-transistors (MOSFETs) is their ability to form a stable oxide SiO$_2$ by thermal oxidation. This allows the development of SiC MOSFETs using techniques similar to Si-based technology, but the formation of the SiO$_2$/SiC interface results in a large number of defect levels in the bandgap. Standard 4H-SiC MOSFETs are fabricated on the Si side of a 4H-SiC wafer (Si face) in dry O$_2$ ambient.\cite{kimoto} On the other hand, for the C side of a 4H-SiC wafer (C face), the MOS fabricated by wet oxidation has a low interface defect density\cite{ApplPhysLett.84.002088,MicroelectronEng.178.000186} and its field-effect mobility easily exceeds 70 cm$^2$V$^{-1}$s$^{-1}$, which is much higher than standard values for Si-face MOSFETs (40 cm$^2$V$^{-1}$s$^{-1}$).\cite{MaterSciForum.600-603.000675,ApplPhysExpress.12.021003} It is generally considered that the interface defects in wet oxidation are related to the field-effect mobility.\cite{ECSTrans.80.000147} Umeda and coworkers have applied electrically detected magnetic resonance (EDMR) spectroscopy to wet-oxidized C faces and found several characteristic defects consisting of carbon-antisite--carbon-vacancy (C$_{\mathrm{Si}}$V$_{\mathrm{C}}$) defects and divacancies (V$_{\mathrm{Si}}$V$_{\mathrm{C}}$).\cite{ApplPhysLett.115.151602,JApplPhys.125.065302,PhysRevB.75.245202,PhysRevLett.96.055501} The C$_{\mathrm{Si}}$V$_{\mathrm{C}}$ and V$_{\mathrm{Si}}$V$_{\mathrm{C}}$ defects are respectively classified as c-axial and basal according to the position and direction of the dangling bonds (DBs) of the C atom. It is reported that C$_{\mathrm{Si}}$V$_{\mathrm{C}}$ c-axial (V$_{\mathrm{Si}}$V$_{\mathrm{C}}$ basal) defects are detected but C$_{\mathrm{Si}}$V$_{\mathrm{C}}$ basal (V$_{\mathrm{Si}}$V$_{\mathrm{C}}$ c-axial) defects are not detected by EDMR spechtroscopy. \cite{ApplPhysLett.115.151602} The reason why only the C$_{\mathrm{Si}}$V$_{\mathrm{C}}$ c-axial and V$_{\mathrm{Si}}$V$_{\mathrm{C}}$ basal defects are detected on the wet-oxidized C face is not clear. In this study, we investigate the stability and formation mechanism of the defects at C face to clarify the characteristic behavior of their generation using first-principles calculations.

The rest of this paper is organized as follows. In Sec.~\ref{sec:Methods}, we state the computational procedure of the density functional theory calculation for the defects at C face. The formation energy and formation mechanism are discussed in Sec.~\ref{sec:Results_and_discussion} and we conclude the study for the characteristic behavior of the defects at C face in Sec.~\ref{sec:Conclusions}.
\section{Methods}
\label{sec:Methods}
Two models are considered in this study: an interface model and a bulk model. Density functional theory\cite{PhysRev.136.B864} calculations are performed using the RSPACE code,\cite{KikujiHirose2005} which is based on a real-space finite-difference approach. \cite{PhysRevLett.72.1240,PhysRevB.50.11355,PhysRevLett.82.5016, PhysRevB.72.085115,PhysRevB.82.205115} The local density approximation parameterized by Vosko {\it et al.} is used as the exchange-correlation functional.\cite{CanJPhys.58.1200} The projecter augmented wave method\cite{PhysRevB.50.17953} is employed for Si, C, and O atoms and the norm-conserving pseudopotential is used for H atoms.\cite{note1,PhysRevLett.48.1425,PhysRevB.43.1993} Supercell sizes are 12.32$\times$10.67$\times$10.06 \AA$^3$ for the bulk model and 12.32$\times$10.67$\times$30.18 \AA$^3$ for the interface model. The numbers of real-space grids are 72$\times$64$\times$60 for the bulk model and 72$\times$64$\times$180 for the interface model. The $\vecvar{k}$-point meshes in the Brillouin zone are 2$\times$2$\times$2 for the bulk model and 2$\times$2$\times$1 for the interface model. In the bulk model, four bilayers in which 16 Si and 16 C atoms are contained in a bilayer are stacked along the (000$\bar{1}$) direction. In the case of the interface model, the interface is represented by a periodic slab consisting of six bilayers of SiC with a vacuum space of 9 \AA \: in the direction perpendicular to the interface. The Si DBs at the bottom layer are terminated by H atoms and the DBs at the top layer are terminated by OH groups. Structural optimization is performed until the maximum Hellmann--Feynman force is less than 0.05 eV/\AA. The atoms in the bottom two layers of the interface model are fixed to simulate an infinitely large solid, while the positions of the other atoms are relaxed.

\begin{figure}
\begin{center}
\includegraphics{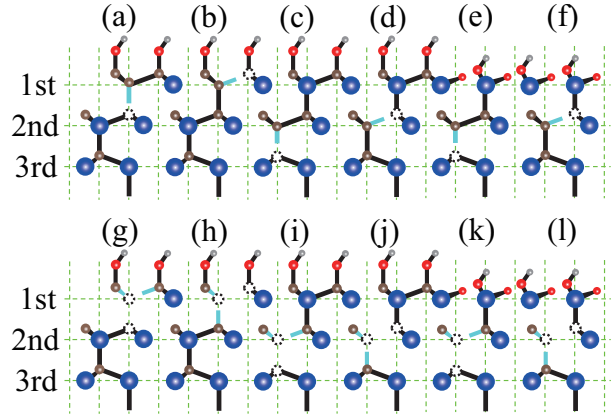}
\caption{Atomic structures of C$_{\mathrm{Si}}$V$_{\mathrm{C}}$ and V$_{\mathrm{Si}}$V$_{\mathrm{C}}$ defects investigated in this study. (a)-(l) represent models A-L, respectively. Blue, brown, red, and gray balls correspond to Si, C, O, and H atoms, respectively. Dotted circles indicate vacancies and blue lines indicate DBs.}
\label{fig:1}
\end{center}
\end{figure}

\begin{table}
\caption{Positions of vacancies and antisites in the models illustrated in Fig.~ \ref{fig:1} and differences in formation energies of defects from that of the most stable defect. In the formation energy, the models whose energies are indicated by hyphens are not stable as mentioned in the text. The formation energies of the most stable configurations of the C$_{\mathrm{Si}}$V$_{\mathrm{C}}$ and V$_{\mathrm{Si}}$V$_{\mathrm{C}}$ defects are set to be zero.}
\begin{tabular}{c|c|cc|cc|cc|c}
\hline \hline
 & & \multicolumn{2}{c|}{1st}& \multicolumn{2}{c|}{2nd}& \multicolumn{2}{c|}{3rd} & \\
Model& Defect type & V$_{\mathrm{C}}$ & C$_{\mathrm{Si}}$(V$_{\mathrm{Si}}$) & V$_{\mathrm{C}}$ & C$_{\mathrm{Si}}$(V$_{\mathrm{Si}}$) & V$_{\mathrm{C}}$ & C$_{\mathrm{Si}}$(V$_{\mathrm{Si}}$) & $\Delta E_{\mathrm{f}}$ (eV) \\
\hline
A & C$_{\mathrm{Si}}$V$_{\mathrm{C}}$ c-axial & 0 & 1 & 1 & 0 & 0 & 0 &  0.00 \\
B & C$_{\mathrm{Si}}$V$_{\mathrm{C}}$ basal   & 1 & 1 & 0 & 0 & 0 & 0 & ---   \\
C & C$_{\mathrm{Si}}$V$_{\mathrm{C}}$ c-axial & 0 & 0 & 0 & 1 & 1 & 0 & +1.73 \\
D & C$_{\mathrm{Si}}$V$_{\mathrm{C}}$ basal   & 0 & 0 & 1 & 1 & 0 & 0 & +0.91 \\
E & C$_{\mathrm{Si}}$V$_{\mathrm{C}}$ c-axial & 0 & 0 & 0 & 1 & 1 & 0 & +2.14 \\
F & C$_{\mathrm{Si}}$V$_{\mathrm{C}}$ basal   & 0 & 0 & 1 & 1 & 0 & 0 & ---   \\
\hline
G & V$_{\mathrm{Si}}$V$_{\mathrm{C}}$ c-axial & 0 & 1 & 1 & 0 & 0 & 0 & ---   \\
H & V$_{\mathrm{Si}}$V$_{\mathrm{C}}$ basal   & 1 & 1 & 0 & 0 & 0 & 0 & ---   \\
I & V$_{\mathrm{Si}}$V$_{\mathrm{C}}$ c-axial & 0 & 0 & 0 & 1 & 1 & 0 & +0.11 \\
J & V$_{\mathrm{Si}}$V$_{\mathrm{C}}$ basal   & 0 & 0 & 1 & 1 & 0 & 0 &  0.00 \\
K & V$_{\mathrm{Si}}$V$_{\mathrm{C}}$ c-axial & 0 & 0 & 0 & 1 & 1 & 0 & +1.02 \\
L & V$_{\mathrm{Si}}$V$_{\mathrm{C}}$ basal   & 0 & 0 & 1 & 1 & 0 & 0 & +0.67 \\
\hline \hline
\end{tabular}
\label{tbl:1}
\end{table}

\section{Results and discussion}
\label{sec:Results_and_discussion}
\begin{figure}
\begin{center}
\includegraphics{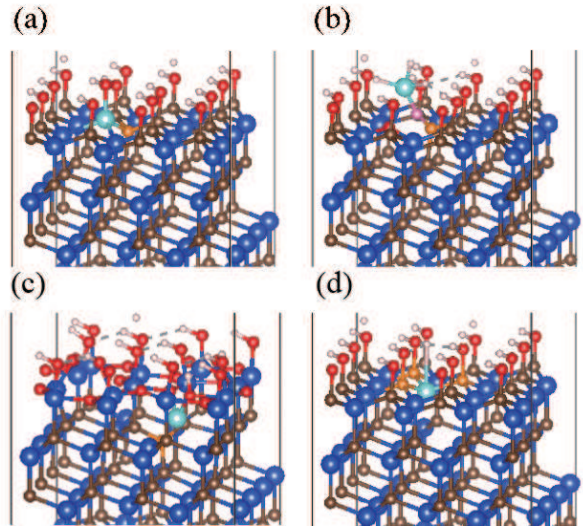}
\caption{Atomic structures of defects where DBs of C atoms are formed in the first bilayer and terminated by O or Si atoms of SiO$_2$. (a) model B terminated by Si atom, (b) model B terminated by O atom, (c) model F terminated by Si atom, and (d) model G terminated by Si atom. Orange, sky blue, and purple balls correspond to C atoms with DBs and Si and O atoms in SiO$_2$, respectively. The meanings of the other symbols are the same as those in Fig.~\ref{fig:1}.}
\label{fig:2}
\end{center}
\end{figure}

The candidate atomic structures of the C$_{\mathrm{Si}}$V$_{\mathrm{C}}$ and V$_{\mathrm{Si}}$V$_{\mathrm{C}}$ defects at the interface are shown in Fig.~\ref{fig:1}. Although considerable theoretical and experimental effort has been devoted to revealing the interface atomic structure of SiC(0001)/SiO$_2$ interface\cite{ApplPhysRev.2.021307,ApplPhysLett.77.2186,ApplPhysLett.93.022108,ApplPhysLett.95.032108,ApplPhysLett.99.021907,ApplPhysLett.97.071908,MaterSciForum.679-680.330,PhysRevB.96.115311}, little is known about the atomic structure of of SiC(000$\bar{1}$)/SiO$_2$ interface. Since there are no experimental studies reporting that the O atoms in SiO$_2$ are always bonded to the C atoms of the C face, we also consider interface models where CO bonds are absent. The atomic structures of the C$_{\mathrm{Si}}$V$_{\mathrm{C}}$ and V$_{\mathrm{Si}}$V$_{\mathrm{C}}$ defects are named as models A to L according to the positions of the DBs relative to the C face. Table~\ref{tbl:1} shows the positions of the vacancies and antisites. In accordance with the net acceptor concentration in Ref.~\citen{ApplPhysLett.115.151602} and the positions of the defect states obtained by theoretical calculations,\cite{PhysRevLett.96.145501,PhysRevB.70.201204,PhysRevB.91.121201} the charged states of the defects are set to $\pm$0 and $-$1 for the C$_{\mathrm{Si}}$V$_{\mathrm{C}}$ and V$_{\mathrm{Si}}$V$_{\mathrm{C}}$ defects, respectively. The atomic structures in which the C atom in the first bilayer (models B, G, and H) and in the second bilayer close to the SiO$_2$ (model F) has DBs are unstable; the DBs of the C atom in the first bilayer are easily terminated by infiltrating from the SiO$_2$ side as illustrated in Fig.~\ref{fig:2}. The last column of Table~\ref{tbl:1} shows the differences in the formation energies of the defects at the interface from the model with the lowest formation energy. The formation energy of the C$_{\mathrm{Si}}$V$_{\mathrm{C}}$ defect is defined as
\begin{equation*}
E_\mathrm{f}=E_\mathrm{w/ \: C_{Si}V_C}+E(\mathrm{Si})-E_\mathrm{interface}
\end{equation*}
with $E_{\mathrm{w/ \: C_{Si}V_C}}$ and $E_{\mathrm{interface}}$ being the total energies of the interface with and without the C$_{\mathrm{Si}}$V$_{\mathrm{C}}$ defect, respectively, and $E(\mathrm{Si})$ is the total energy of a Si atom in bulk. The difference in the formation energy of the $\mathrm{V_{Si}V_C}$ defects is calculated in a similar manner. The C$_{\mathrm{Si}}$V$_{\mathrm{C}}$ defect is considered to be formed by the migration of the neighboring C atom to the V$_{\mathrm{Si}}$ site after the formation of the silicon vacancy. It is found that the formation energy of the C$_{\mathrm{Si}}$V$_{\mathrm{C}}$ c-axial defects associated with the silicon vacancy at the first bilayer is the lowest. If the silicon vacancy was generated at the second bilayer, the C$_{\mathrm{Si}}$V$_{\mathrm{C}}$ basal defect (model D) would be preferentally formed because the formation energy of model D is lower than that of model C. However, since the formation energy of the silicon vacancy in the first bilayer is 1.8 eV lower than that in the second bilayer, hardly any silicon vacancies are generated in the second bilayer. This agrees with the fact that the density of C$_{\mathrm{Si}}$V$_{\mathrm{C}}$ c-axial defects detected in the previous experimental study was much higher than that of other defects.\cite{ApplPhysLett.115.151602}

The formation energy of the V$_{\mathrm{Si}}$V$_{\mathrm{C}}$ basal defect is 0.11 eV lower than that of the c-axial defect, which is a rather small difference. The C atom at the carbon antisite could be removed from the C$_{\mathrm{Si}}$V$_{\mathrm{C}}$ defect, giving rise to the V$_{\mathrm{Si}}$V$_{\mathrm{C}}$ defect. However, the process in which V$_{\mathrm{Si}}$V$_{\mathrm{C}}$ basal defects appear after the formation of  C$_{\mathrm{Si}}$V$_{\mathrm{C}}$ basal defects is expected to hardly occur because the existence of  C$_{\mathrm{Si}}$V$_{\mathrm{C}}$ basal defects is not suggested by both the EDMR measurement and the above-mentioned discussion of C$_{\mathrm{Si}}$V$_{\mathrm{C}}$ defects. In the process where the Si atom is desorbed after the formation of the carbon vacancy, the formation energy of the V$_{\mathrm{C}}$ defect at the basal site is 0.28 eV lower than that at the c-axial defect. The formation energy of the silicon vacancy in the second bilayer in the presence of the carbon vacancy at the basal site is 3.59 eV lower than that in the absence of the carbon vacancy because four DBs are formed by the desorption of the Si atom, whereas the number of DBs is increased by only two by the removal of the Si atom associated with the carbon vacancy. According to the energy difference between the basal and c-axial defects as well as the difference between the intermediate structures of the carbon vacancies, our results are consistent with the experimental findings that carbon vacancies are detected characteristically in the C face and V$_{\mathrm{Si}}$V$_{\mathrm{C}}$ basal defects are more preferentially generated than c-axial defects.\cite{ApplPhysLett.115.151602} Thus, the positions of the vacancies and antisites shown in Table \ref{tbl:1} are relevant to the preference of the generation of the C$_{\mathrm{Si}}$V$_{\mathrm{C}}$ c-axial (V$_{\mathrm{Si}}$V$_{\mathrm{C}}$ basal) defect over C$_{\mathrm{Si}}$V$_{\mathrm{C}}$ basal (V$_{\mathrm{Si}}$V$_{\mathrm{C}}$ c-axial) defect because defects are more easily formed at the interface than in the bulk.

\begin{table}
\caption{Magnetic moments of defects.}
\label{tbl:3}
\begin{tabular}{c|cc}
\hline \hline
Model & Bulk ($\mu_{\mathrm{B}}$) & Interface ($\mu_{\mathrm{B}}$) \\
\hline
C$_{\mathrm{Si}}$V$_{\mathrm{C}}(+)$ c-axial & 1 & 1 \\
V$_{\mathrm{Si}}$V$_{\mathrm{C}}$ basal & 2 & 2 \\
\hline \hline
\end{tabular}
\end{table}

\begin{figure}
\begin{center}
\includegraphics{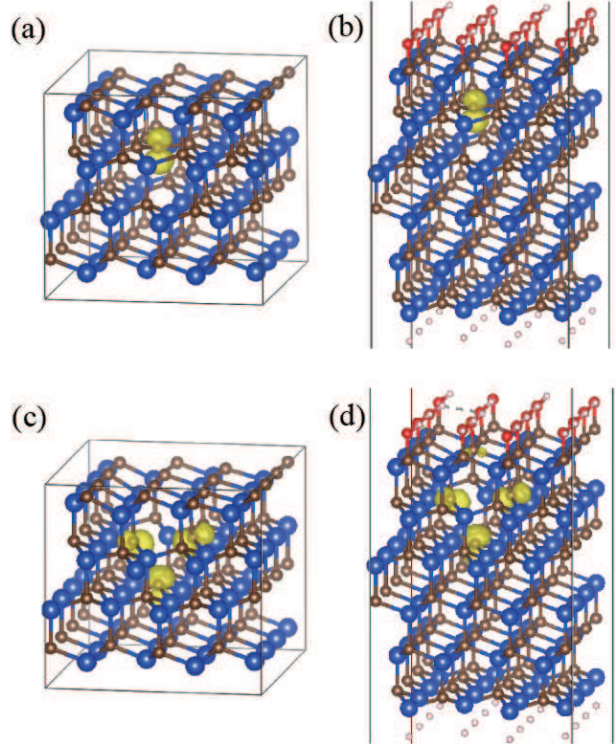}
\caption{Difference in spatial distribution of electron density between up and down spins. (a) C$_{\mathrm{Si}}$V$_{\mathrm{C}}$ c-axial defect in bulk, (b) C$_{\mathrm{Si}}$V$_{\mathrm{C}}$ c-axial defect at interface, (c) V$_{\mathrm{Si}}$V$_{\mathrm{C}}$ basal defect in bulk, and (d) V$_{\mathrm{Si}}$V$_{\mathrm{C}}$ basal defect at interface. The yellow region is the isosurface of the distribution. The meanings of the other symbols are the same as those in Fig.~\ref{fig:1}.}
\label{fig:3}
\end{center}
\end{figure}

\begin{figure*}
\begin{center}
\includegraphics{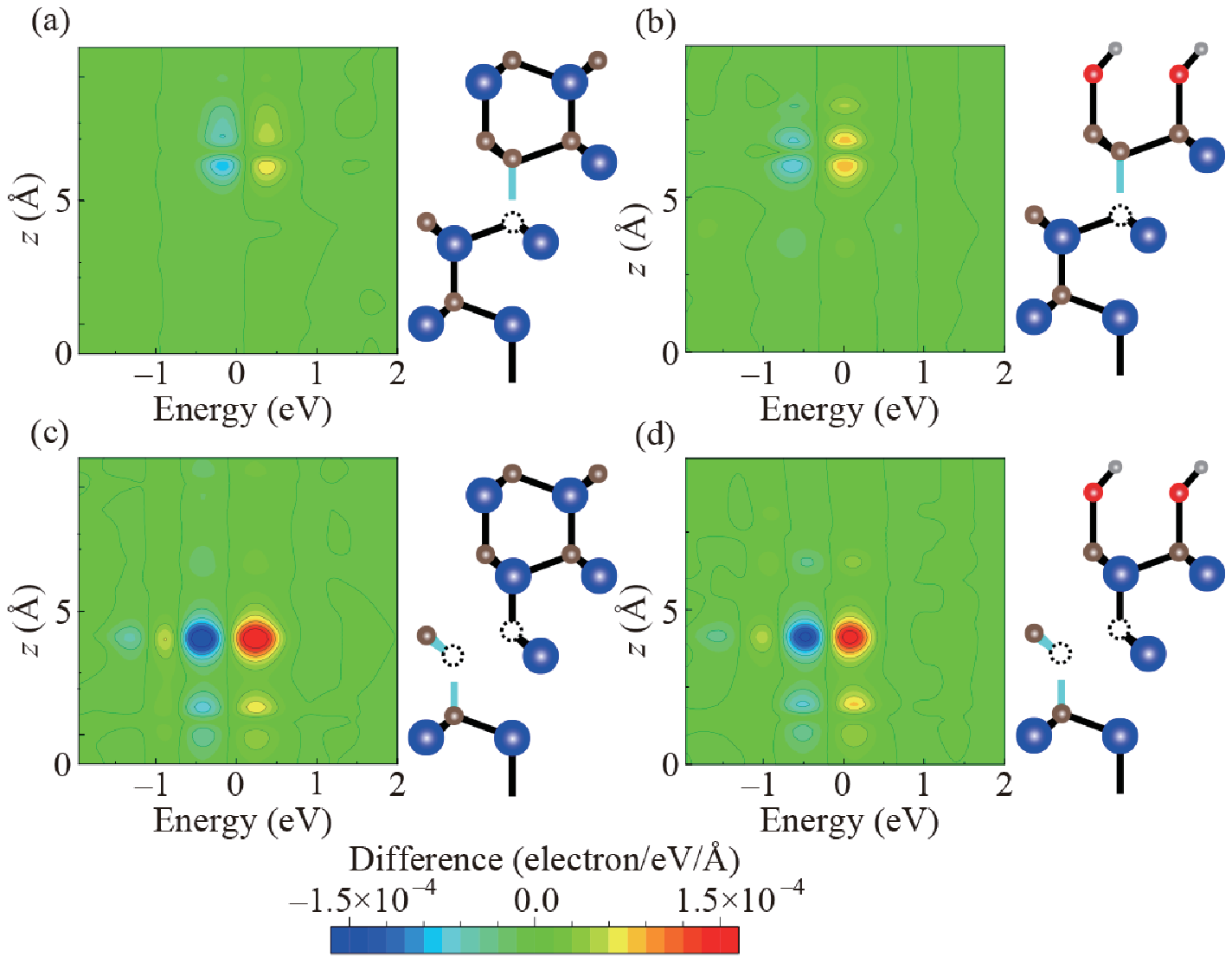}
\caption{Difference in local density of states between up- and down-spin electrons. (a) C$_{\mathrm{Si}}$V$_{\mathrm{C}}$ c-axial defect in bulk, (b) C$_{\mathrm{Si}}$V$_{\mathrm{C}}$ c-axial defect at interface, (c) V$_{\mathrm{Si}}$V$_{\mathrm{C}}$ basal defect in bulk, and (d) V$_{\mathrm{Si}}$V$_{\mathrm{C}}$ basal defect at interface. Atomic structures are illustrated as a visual aid. The meanings of the symbols are the same as those in Fig.~\ref{fig:1}.}
\label{fig:4}
\end{center}
\end{figure*}

The magnetic moments of the C$_{\mathrm{Si}}$V$_{\mathrm{C}}$(+1) c-axial and V$_{\mathrm{Si}}$V$_{\mathrm{C}}$($\pm$0) basal defects both in the bulk and at the interface are shown in Table~\ref{tbl:3}, in which one electron is removed from the defects to imitate the EDMR observations. Although the C$_{\mathrm{Si}}$V$_{\mathrm{C}}$($\pm$0) defects are not spin-polarized, they are spin-polarized by removing one electron. The magnetic moments in the bulk agree with those at the interface. Figure~\ref{fig:3} shows the difference in the spatial distribution of electron density between up and down spins. The DBs of the C atoms are spin-polarized in the bulk and at the interface, and their spatial distributions are similar. The difference in the local density of states (LDOS) between up and down spins is shown in Fig.~\ref{fig:4}. The LDOS is calculated as
\begin{equation*}
\rho(z,E)=\sum_{i,k} \int |\Psi_{i,k}(x,y,z)|^2 \mathrm{d}x \mathrm{d}y \times N \e^{-\alpha(E-\varepsilon_{i,k})^2},
\end{equation*}
where $\varepsilon_{i,k}$ are the eigenvalues of the wavefunction, with indexes $i$ and $k$ denoting the eigenstate and the k-point respectively. $N(=2\sqrt{\frac{\pi}{\alpha}})$ is the normalization factor, where $\alpha$ is the smearing factor, here set to 13.5~eV$^{-2}$. The LDOS at the interface is consistent with that in the bulk. The spatial distributions of electron density and LDOS indicate that the proposed interface defect structure can be detected as the C$_{\mathrm{Si}}$V$_{\mathrm{C}}$ c-axial and V$_{\mathrm{Si}}$V$_{\mathrm{C}}$ basal defects by EDMR measurement.

Next, the formation energies of silicon and carbon vacancies are compared with the dry and wet oxidations because the silicon and carbon vacancies are the source of the C$_{\mathrm{Si}}$V$_{\mathrm{C}}$ and V$_{\mathrm{Si}}$V$_{\mathrm{C}}$ defects. The DBs of the wet oxidized interface are passivated by H atoms when they are generated because the hydrogen passivation of the defects is broken by the $\gamma$-ray irradiation before the EDMR measurement.\cite{ApplPhysLett.115.151602} Therefore, the models for the wet oxidation contain the hydrogen passivations of DBs of C atoms. However, the DBs of Si atoms are not terminated because the hydrogen passivation in the SiC bulk is found to be not stable. The formation of silicon vacancies by dry oxidation is
\begin{equation*}
\mathrm{Si_n}\mathrm{C_n}+\mathrm{O_2} \rightarrow \mathrm{Si}_{\mathrm{n}-1}\mathrm{C_n}+\mathrm{SiO_2}
\end{equation*}
and that by wet oxidation is
\begin{equation*}
\mathrm{Si_n}\mathrm{C_n}+2\mathrm{H_2O} \rightarrow \mathrm{Si}_{\mathrm{n}-1}\mathrm{C_nH_4}+\mathrm{SiO_2}.
\end{equation*}
Since C-H bonds are more stable than H-H bonds, the DBs in the SiC substrate are preferentially terminated by H atoms. The formation energies for the dry and wet oxidations are
\begin{equation*}
E_{\mathrm{form}} = \left( E_{\mathrm{interface}} +2 \mu_{\mathrm{O}} \right)- \left( E(\mathrm{V}_{\mathrm{Si}}) + E(\mathrm{SiO_2}) \right) 
\end{equation*}
and
\begin{equation*}
E_{\mathrm{form}} = \left( E_{\mathrm{interface}} + 2 E(\mathrm{H_2O}) \right)- \left( E(\mathrm{V_{Si}H_4}) + E(\mathrm{SiO_2}) \right),
\end{equation*}
respectively. Here, $E_{\mathrm{interface}}$, $E(\mathrm{V_{Si}})$, $E(\mathrm{V_{Si}H_4})$, and $E(\mathrm{SiO_2})$ are the total energies of the interface without any defects, the interface with the silicon vacancy at the first bilayer, the interface in which the DBs neighboring the silicon vacancy are terminated by H atoms, and the total energy of a SiO$_2$ unit in $\alpha$-quartz SiO$_2$, respectively. $\mu_{\mathrm{O}}$ is the chemical potential of an O atom. In the case of O-atom rich condition, the chemical potential of an O atom is derived from the total energy of an isolated oxygen molecule, $E(\mathrm{O_2})$, as
\begin{equation*}
\mu_{\mathrm{O}}^{\mathrm{r}} = E(\mathrm{O_2})/2.
\end{equation*}
At the oxidation front, since O atoms are hardly provided, which is O-atom poor condition, the chemical potential of an O atom is expressed as
\begin{equation*}
\mu_{\mathrm{O}}^{\mathrm{p}} = \left( E(\mathrm{SiO_2}) - E(\mathrm{Si}) \right)/2.
\end{equation*}
The difference in the chemical potential of O atom between the rich and poor conditions, $\mu_{\mathrm{O}}^{\mathrm{r}}-\mu_{\mathrm{O}}^{\mathrm{p}}$, is 4.97 eV. The formation energies of the dry and wet oxidations are $-$6.50(3.44) and 3.98(3.98) eV, respectively, for $\mu_{\mathrm{O}}=\mu_{\mathrm{O}}^{\mathrm{p}}$ ($\mu_{\mathrm{O}}=\mu_{\mathrm{O}}^{\mathrm{r}}$).
 As in the case of the carbon vacancy, the reaction in dry oxidation is
\begin{equation*}
\mathrm{Si_nC_n}+\frac{1}{2}\mathrm{O_2} \rightarrow \mathrm{Si_nC}_{\mathrm{n}-1}+\mathrm{CO}
\end{equation*}
and that in wet oxidation is
\begin{equation*}
\mathrm{Si_nC_n}+2\mathrm{H_2O} \rightarrow \mathrm{Si_nC}_{\mathrm{n}-1}+\mathrm{CH_4}+\mathrm{O_2}.
\end{equation*}
The formation energies in the dry and wet oxidations are
\begin{equation*}
E_{\mathrm{form}} = \left( E_{\mathrm{interface}} + \mu_{\mathrm{O}} \right)- \left( E(\mathrm{V_C}) + E(\mathrm{CO}) \right) 
\end{equation*}
with an energy gain of $-$8.87($-$3.90) eV for $\mu_{\mathrm{O}}=\mu_{\mathrm{O}}^{\mathrm{p}}$ ($\mu_{\mathrm{O}}=\mu_{\mathrm{O}}^{\mathrm{r}}$) and 
\begin{equation*}
E_{\mathrm{form}} = \left( E_{\mathrm{interface}} + 2 E(\mathrm{H_2O}) \right)- \left( E(\mathrm{V_C}) + E(\mathrm{CH_4}) + 2 \mu_{\mathrm{O}} \right)
\end{equation*}
with an energy gain of 1.40($-$8.54) eV for $\mu_{\mathrm{O}}=\mu_{\mathrm{O}}^{\mathrm{p}}$ ($\mu_{\mathrm{O}}=\mu_{\mathrm{O}}^{\mathrm{r}}$), respectively. In addition, when an O atom is assumed to be taken from SiO$_2$ with an oxygen vacancy, the difference in the chemical potential from that of an oxygen molecule, $\mu_{\mathrm{O}}^{\mathrm{r}}-\mu_{\mathrm{O}}$, is larger than 5 eV,\cite{JpnJApplPhys.37.L232,AIPAdv.7.015309} which is further O-atom poor condition. Thus, the silicon and carbon vacancies are more easily generated by wet oxidation than by dry oxidation near the SiC/SiO$_2$ interface because the formation energy of the vacancies by wet oxidation is larger than that by the dry oxidation under the O-atom poor condition ($\mu_{\mathrm{O}} \approx \mu_{\mathrm{O}}^{\mathrm{p}}$). From the above results, we conclude that the characteristic behavior of the formation of the C$_{\mathrm{Si}}$V$_{\mathrm{C}}$ c-axial and V$_{\mathrm{Si}}$V$_{\mathrm{C}}$ basal defects by wet oxidation is induced by the termination of DBs around the vacancies. 

\section{Conclusions}
\label{sec:Conclusions}
We have proposed the atomic structures of the C$_{\mathrm{Si}}$V$_{\mathrm{C}}$ and V$_{\mathrm{Si}}$V$_{\mathrm{C}}$ defects and the mechanism of their generation. In the case of the C$_{\mathrm{Si}}$V$_{\mathrm{C}}$ defect, the DBs of the C$_{\mathrm{Si}}$V$_{\mathrm{C}}$ basal defect formed after the generation of the silicon vacancy in the first bilayer are easily terminated by O or Si atoms in SiO$_2$. The formation energy of the C$_{\mathrm{Si}}$V$_{\mathrm{C}}$ c-axial defect generated after the formation of the silicon vacancy in the first bilayer is lower than that of the C$_{\mathrm{Si}}$V$_{\mathrm{C}}$ basal defect after the formation of the silicon vacancy in the second bilayer, indicating that the C$_{\mathrm{Si}}$V$_{\mathrm{C}}$ c-axial defect is  preferentially generated. In the case of the V$_{\mathrm{Si}}$V$_{\mathrm{C}}$ defect, the DBs of the V$_{\mathrm{Si}}$V$_{\mathrm{C}}$ defects in the first bilayer are also terminated by O or Si atoms in SiO$_2$. The formation energy of the carbon vacancy at the basal site is lower than that at the c-axial site. The formation energy of the silicon vacancy in the second bilayer in the presence of the carbon vacancy at the basal site is lower than that in the case of no carbon vacancies. The V$_{\mathrm{Si}}$V$_{\mathrm{C}}$ basal defect is more preferentially generated than the V$_{\mathrm{Si}}$V$_{\mathrm{C}}$ c-axial defect. In this mechanism, hardly any C$_{\mathrm{Si}}$V$_{\mathrm{C}}$ basal and V$_{\mathrm{Si}}$V$_{\mathrm{C}}$ c-axial defects are generated. These results are consistent with the experimental results of a previous study.\cite{ApplPhysLett.115.151602} Furthermore, the electronic structures of the C$_{\mathrm{Si}}$V$_{\mathrm{C}}$ c-axial and V$_{\mathrm{Si}}$V$_{\mathrm{C}}$ basal defects at the interface are the same as those in the bulk, indicating that our proposed structures correspond to those observed in the experiment. The characteristic property of the generation of C face defects is explained by the positions of the vacancies and antisites in the SiC(000$\bar{1}$) substrate and the termination of DBs around the vacancies. It is fair to conclude that the preferential generation of the C$_{\mathrm{Si}}$V$_{\mathrm{C}}$ c-axial and V$_{\mathrm{Si}}$V$_{\mathrm{C}}$ basal defects are related to the position of the defects along the direction perpendicular to the interface while the generation of the C$_{\mathrm{Si}}$V$_{\mathrm{C}}$ and V$_{\mathrm{Si}}$V$_{\mathrm{C}}$ defects by wet oxidation is relevant to the termination of the DBs.

\acknowledgments 
This work was partially financially supported by MEXT as part of the ``Program for Promoting Researches on the Supercomputer Fugaku'' (Quantum-Theory-Based Multiscale Simulations toward the Development of Next-Generation Energy-Saving Semiconductor Devices, JPMXP1020200205), JSPS KAKENHI (JP16H03865), Kurata Grants, and the Iwatani Naoji Foundation. The numerical calculations were carried out using the computer facilities of the Institute for Solid State Physics at The University of Tokyo, the Center for Computational Sciences at University of Tsukuba, and the supercomputer Fugaku provided by the RIKEN Center for Computational Science (Project ID: hp210170).

\end{document}